# The selforganization phenomenon in RayleighBenard convection :a hydrodynamics analysis

Zhe Wu

*Department of Physics and Key Laboratory of Micro-nano Measurement-Manipulation and Physics (Ministry of Education) Beijing University of Aeronautics and Astronautics, Beijing 100191, China*

## 1.Abstract

The evolution of three-dimensional, cellular convective flows in a plane horizontal layer of a Boussinesq fluid heated from below is well studied. Here we review results from the investigation of this system as well as a number of related and novel numerical findings. We present theoretical results for pattern formation in Rayleigh-Be ́nard convection with solving NS equation under the Boussinesq approximation . System of equations were reduced to 2-dimension form to simplify the analysis.The evolution of the flow agree with the idea of the flow achieving an optimal form.

## 2.Background

Symmetry plays a crucial role in selecting the patterns displayed in physical, chemical, and biological systems as they are driven away from equilibrium [1].A typical form of convection in a plane horizontal fluid layer is Quasi-two-dimensional roll flows below Rayleigh-Benard convection over a widerange of Rayleigh numbers R.This flow provides that the spatial distribution of the material parameters of the fluid does not exhibit any significant up-down asymmetry with respect to the horizontal midplane of the layer[2]

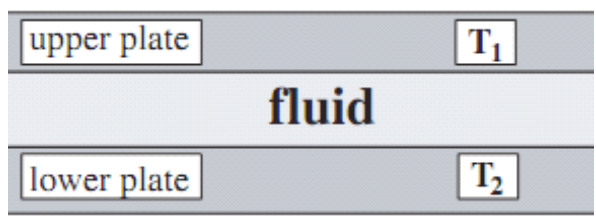

Figure 1. Fluid motion in the layer of depth d is driven both by the imposed temperature difference T = T2−T1

The flow of water communicating between the two systems is reflected in the role of symmetry pattern . In the parameter region as set up above the behaviour of exponentially growing linear modes is captured by equations for the spectral mode amplitudes A, whose terms must satisfy spatial and temporal symmetries.[3]

## 3.Theoretical analysis method:hypothesis and equations

1.Boussinesq approximation
The approximation is extremely accurate for many such flows, and makes the mathematics and physics simpler.The Boussinesq approximation is applicable because when considering a flow of warm and cold water of density ρ1 and ρ2 one needs only consider a single density ρ: the difference Δρ =ρ1 −ρ2 is negligible.Boussinesq flows are not rare in nature, industry, and the built environment[4]. Upside-down feature of Boussinesq flows will be obtained provided that the identities of the fluids are reversed[5]. The Boussinesq approximation will be inaccurate once the  density difference Δρ /ρ is of order unity. A Typical example of a non-Boussinesq flow is bubbles rising in water. Difference is very obvious between air bubbles rising in water and the behaviour of water falling in air. Related experiments was made by Duineveld (1995), who apply hyper-clean water to study rising air bubbles.

2.Navier–Stokes equations for the Newtonian fluids

$$\frac{dV}{dt} = F - \frac{1}{\rho}\nabla P + \frac{2}{3}\frac{\mu}{\rho}\nabla \cdot \nabla V + \frac{\mu}{\rho}\nabla^2 \qquad 1$$

The third item above is 0. The equation can be reduced to

$$\frac{dV}{dt} = F - \frac{1}{\rho}\nabla P + \gamma\nabla^2 V \qquad 2.$$

$\gamma = \frac{\mu}{\rho}$ is the viscosity coefficients. $\frac{dV}{dt}$ is the acceleration for units of fluid. F is

the force unit fluid tolerate. The second item is pressure gradient. The last item is the viscous force. If the fluids experience a bulk movement,the viscous force turned to zero.[7]

Equation 2 can be written in Cartesian coordinates as follows:

$$\frac{\partial u}{\partial t} + u\frac{\partial u}{\partial x} + v\frac{\partial u}{\partial y} + \omega\frac{\partial u}{\partial z} = F_x - \frac{1}{\rho}\frac{\partial P}{\partial x} + \gamma\nabla^2 u \qquad 3$$

$$\frac{\partial v}{\partial t} + u\frac{\partial v}{\partial x} + v\frac{\partial v}{\partial y} + \omega\frac{\partial v}{\partial z} = F_y - \frac{1}{\rho}\frac{\partial P}{\partial y} + \gamma\nabla^2 v \qquad 4$$

$$\frac{\partial \omega}{\partial t} + u\frac{\partial \omega}{\partial x} + v\frac{\partial \omega}{\partial y} + \omega\frac{\partial \omega}{\partial z} = F_z - \frac{1}{\rho}\frac{\partial P}{\partial z} + \gamma\nabla^2 \omega \qquad 5$$

3.set of heat convection equations

$$\frac{dV}{dt} = -\frac{1}{\rho}\nabla P - g\alpha T' + \gamma \nabla^2 V \qquad 6$$

$$\nabla \cdot \overline{\rho} V = 0 \qquad 7$$

$$\frac{dT'}{dt} = \frac{\partial T'}{\partial t} + V \cdot \nabla T' = \chi_1 \nabla^2 T' \qquad 8$$

In the equation 8 $\chi_1$ is the heat diffusion coefficient, $\alpha T'$ is the heating effect coefficient in equation 6. From the equations above we can obtain that heating effect $\chi_1 \nabla^2 T'$ causes temperature field T'. This will drive fluid through buoyancy item $-g\alpha T'$. Immediately after the flows the heating energy will be transfered to change the distribution of the former temperature. This kind of inhibition is crucial in our later analysis.

## 4.Analysis for the RayleighBenard convection

1.Description of Benard convection
    Benard convection is naturally exist in the convection in thin shell fluids.As an envolution with time a regular pattern of convection cells known as Bénard cells will appear. For then on the convection patterns are the example of self-organizing nonlinear systems.[8]
Hypothesis are established as follows:a thin liquid is lie in between two planes.The thickness of the liquid is H which is far less than the width of the planes. Only under this condition,we can ignore the edge influence. At the beginning, the liquid will reach thermal equilibrium and become balanced with the external environment. T1 and T2 represents the temperature of upper and lower planes(Fig 1). Before the heating process, the temperature difference between two planes are zero. Heating from the lower plane,then the T2 will be higher than T1. The system equilibrium will be broken. When the constraint is at a low point, accordingly the ΔT is very small, the lower plane will transfer energy by heat conduction. The temperature distribution in the upper layer of the fluid and the density distribution will in the linear form.T=T2-Zγ; ρ=ρ2(1-η(T-T2)). The η and γ are scale coefficient. With the continuing heating for the lower plane, the system will leave the equilibrium much further. When the ΔT reaches a critical point, the upper-lower fluid can be observed[9]. After experience an irregular phase,the whole fluid layer will separate into many regular units,the fluid will show small sprays. These sprays is well localized and appear to be the hexagon.[Fig 2] The fluid in the closed units will rise up in the center and descend along the border. The array of the units will be obeyed as the left and right spin with the neighbours. Continuing increase the temperature,the closed units will become unsteady thus the chaos will be dominant.[10]

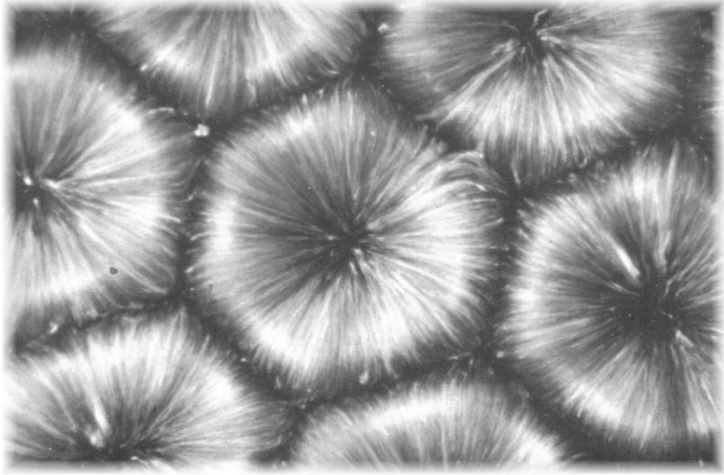

Figure 2. perfect two-vortex cell

2.The linear simplification for the RayleighBenard convection

Leave out the nonlinear items in the heat convection equations, together with the equations 3,4,5 we can obtain the linearized equations as follows:

$$\begin{cases} \dfrac{\partial u}{\partial t} = -\dfrac{1}{\rho_0}\dfrac{\partial P'}{\partial x} + \gamma\nabla^2 u \cdots\cdots\cdots\cdots\cdots 9 \\ \dfrac{\partial v}{\partial t} = -\dfrac{1}{\rho_0}\dfrac{\partial P'}{\partial y} + \gamma\nabla^2 v \cdots\cdots\cdots\cdots\cdots 10 \\ \dfrac{\partial \omega}{\partial t} = -\dfrac{1}{\rho_0}\dfrac{\partial P'}{\partial z} + \gamma\nabla^2 \omega + g\alpha\nabla_1^2 T' \cdots\cdots 11 \end{cases}$$

$$\left(\dfrac{\partial}{\partial t} - \chi_1\nabla^2\right)T' = -r\omega \qquad 12$$

$$\dfrac{\partial \rho_0 u}{\partial x} + \dfrac{\partial \rho_0 v}{\partial x} + \dfrac{\partial \rho_0 \omega}{\partial x} = 0 \qquad 13$$

The V and $\chi_1$ can be set up as constant as well as $\rho_0$. From the equations 9,10,11. We can reduce u,v,P'. Two convection equations with two unknown variables can be obtained as follows:

$$\begin{cases} \left(\dfrac{\partial}{\partial t} - v\nabla^2\right)\nabla^2\omega = g\alpha\nabla_1^2 T' \cdots\cdots\cdots 14 \\ \left(\dfrac{\partial}{\partial t} - \chi_1\nabla^2\right)T' = -\gamma\omega \cdots\cdots\cdots\cdots 15 \end{cases}$$

In the equations above $\nabla_1^2 = \dfrac{\partial^2}{\partial x^2} + \dfrac{\partial^2}{\partial y^2}$, form the equations 14,15 we can obtain that heating when T'>0 leads to convection ($\dfrac{\partial \omega}{\partial t} \neq 0$). The results of the convection turned out to change the former heating distribution ($\dfrac{\partial T'}{\partial t} \neq 0$). The viscous property (v) and heat diffusion ($\chi_1$) impede the development of the convection. As the equations 9.10.11 are the forced source diffusion equations and for the $\omega$ 和 T' diffusion equations will be forced source to each other, in this way, the Benard convection is formed.

We locate the origin of coordinate in the thin fluid layer, then the boundary conditions in the top layer and bottom layer can be written as Z=±0.5H; $\dfrac{\partial \omega}{\partial z} = 0$ ; $\omega$ =0.

Assume that the two plane layers will easier for heat transfer than fluid. Introduce the dimensionless time variable $\tau$ and spatial variables x, y, z.

$\tau = \dfrac{t\chi_1}{H^2}$, (x，y，z)=(x，y，z)/H

The solution form for the equation set(9,10,11) can be written as follows:

$$\left.\begin{aligned} \omega(x,y,z,\tau) &= \dfrac{\chi_1}{H}W(z) \\ T'(x,y,z,\tau) &= \gamma H\theta(z) \end{aligned}\right\} e^{\sigma\tau} f(x,y)$$

The item $\chi_1/H$ is the parameter which has the same dimension with the velocity. The item $\gamma H$ is the parameter which has the same dimension with the temperature. H is the thickness of the fluid layer. The Items θ(z), w(z) are dimensionless function. $e^{\sigma\tau}$ is the time growth factor. The f(x,y) is the horizontal distribution of heat convection. It obeys the equation $\frac{\partial^3 f}{\partial x^2}+\frac{\partial^2 f}{\partial y^2}+a^2 f=0$.

Item a is a dimentionless wave number which is in an inverse ratio to the the scale of the convection units. Substitute the solution to the equations 14,15. We can get the equations for the θ(z), w(z).

$$\begin{cases} \sigma-\left(\dfrac{d^2}{dz^2}-a^2\right)\theta = W \\ \left[\dfrac{\sigma}{p_r}-\left(\dfrac{d^2}{dz^2}-a^2\right)\right]\left(\dfrac{d^2}{dz^2}-a^2\right)W = R_a a^2 \theta \end{cases}$$

## 5.Conclusions

In the equations above the Pr= ɣ / x 1 are prandtl number, Ra is rayleigh number. The corresponding boundary conditions listed above can be used here. In this way, we can solve the equations to get θ(z), w(z) with respect to z in principle. Then the heat convection traits can be obtained from $\frac{\partial^3 f}{\partial x^2}+\frac{\partial^2 f}{\partial y^2}+a^2 f=0$.

The simplest form of f can be written as f(x*y)=sin(mx)*sin(ny)[fig3 and fig4]. In the function $m^2+n^2=a$. It corresponds to a rectangular convection units. When x=0,y=0, f(x,y)=0; x=± π /m, y=± π /n, f(x*y)=0.

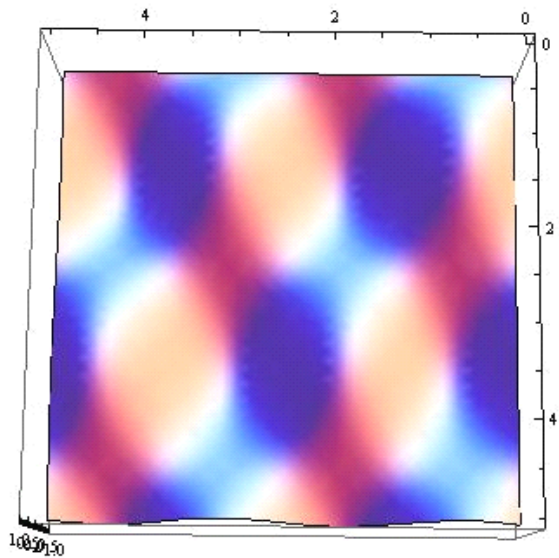

Fig3 a overlook of the distribution with m=2.5,a=4.24

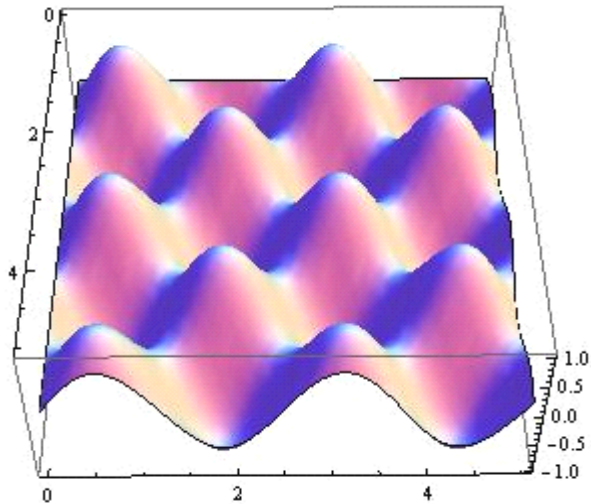

Fig4 a lateral view of the distribution with m=2,a=2.5

## 6.References

[1] M. C. Cross and P. C. Hohenberg, Rev. Mod. Phys. 65, 851(1993)
[2]PHYSICAL REVIEW E 67, 046313 (2003)
[3]Pattern formation in vertically oscillated convection
[4]The validity of the Boussinesq approximationfor liquids and gases
[5]Long-wavelength surface-tension-driven Bénard convection: experimentand theory
[6]The motion of high-Reynolds-number bubblesin inhomogeneous flows
[7]Polyanin, A.D.; Kutepov, A.M.; Vyazmin, A.V.; Kazenin, D.A. (2002), Hydrodynamics, Mass and Heat Transfer in Chemical Engineering, Taylor & Francis, London, ISBN 0-415-27237-8
[8]Rayleigh-Bénard Convection: Structures and Dynamics, Alexander V. Getling, World Scientific Publishing
[9]Visualization of convection loops due to Rayleigh–Benard convection during solidification
[10]Spiral defect chaosin large aspect ratio Rayleigh-Bénard convection